\def\hh{\hskip10pt}
\documentstyle[aps,preprint]{revtex}
\input epsf.tex
\input epsf.sty
\newcommand{\RG}{renormalization group}
\newcommand{\eps}{\varepsilon}

\begin{document}

\preprint{ PM94-19 \\ \hspace*{\parindent}hep-th ymnnn}

\setcounter{figure}{0}
\parskip 10pt
\vfill\eject

\title{Variational solution of the Gross-Neveu model
       \\I.  The large-$N$ limit }

\author{C. Arvanitis\thanks
        {c.arvanitis@ic.ac.uk. Address after
Sept. $1^{\rm st}$ 1994 : Physics Department,
\newline \hspace*{\parindent}
 Imperial College, London.}
F. Geniet\thanks{geniet@lpm.univ-montp2.fr}~
 and A. Neveu\thanks
        {neveu@lpm.univ-montp2.fr.
On sabbatical leave after Sept. $1^{\rm st}$1994
 \newline
        \hspace*{\parindent} at the Physics Department
and Lawrence Berkeley Laboratory,
University of\newline\hspace*{\parindent}
 California, Berkeley, neveu@theorm.lbl.gov}
     \\ Laboratoire de Physique Math\'ematique\thanks
        {Laboratoire associ\'e au Centre
National de la Recherche Scientifique.}
     \\ Universit\'e de Montpellier II-CNRS
     \\ 34095 Montpellier Cedex 05}

\date{March 26, 1995}

\maketitle

\newpage

\begin{abstract}

In this first paper we begin the
application of variational methods
 to renormalizable asymptotically
free field theories, using the Gross-Neveu model
 as a laboratory. This variational method has been
shown to lead to a numerically convergent sequence
 of approximations for the anharmonic oscillator.
Here we perform a sample calculation in lowest
orders, which show the {\em superficially}
disastrous situation of variational calculations
 in quantum field theory, and how in the large-$N$
limit all difficulties go away, as a warm up
exercise for the finite-$N$ case.

\end{abstract}

\newpage

\section{Introduction.} \label{intro}

Many non-perturbative methods have been
developed  in recent years, trying to
improve the asymptotic Taylor expansions
 (in terms of some coupling constant
parameter)
one usually encounters in quantum field theories.
A series of elegant papers in the seventies has
explored and clarified various
aspects
of the large order behavior of perturbation
theory in quantum mechanical and
quantum field theoretical systems.
In some cases this newly acquired knowledge was
successfully used to obtain more
accurate results from the asymptotic perturbative
series via  Borel transforms.
In other cases, especially in four-dimensional field
theoretic models, the question
of what the sum of a badly-behaved perturbation
series means, remains unanswered ~\cite{GUIL}.

Recently, there has been a considerable
interest in variational-type expansions
(\cite{CAS}-\cite{SOLO}).
Because of its conceptual simplicity
and  usually quite astonishing
numerical accuracy, this type of expansion
would be of  particular importance
in quantum field theoretical calculations.
Despite the distinctly intuitive nature of
such a procedure, there has
recently appeared
strong evidence that optimized perturbation
 theory may indeed lead to a
rigorously
convergent series of approximants even in
strong coupling cases.
In particular, the convergence of this
variational-like procedure has been
rigorously established in the case of
"zero" and "one" dimensional field
theories ~\cite{JONE}.

In this paper, we begin the application to quantum
field theory of the
variational improvement of perturbation theory
as described in \cite{p1}. In the case
of the anharmonic oscillator, this method
gives a sequence of approximations which
converges to correct answers as described in
details in \cite{p2,p3}. Here, as a
laboratory, using dimensional regularization,
we study
the $O(2N)$ Gross-Neveu model \cite{p4} in
 $2+\eps$ dimensions. Exact results
 are known in 2 dimensions for the mass gap
 \cite{p5} and the vacuum energy density
 \cite{p6} in terms of $\Lambda_{\overline{MS}}$
for all $N$.

Before comparing variational calculations to
 these exact results, one must first
understand how to handle infinities at $\eps
\rightarrow 0$ in a way compatible
with perturbative renormalization, leaving only
finite physical quantities expressed only
in terms of finite physical renormalization group
invariants of the effective
variational approximation. This is the only
sensible procedure if we want the
physical characteristics of the theory, which
 have led to the specific choice
of the effective variational approximation, to
 survive renormalization.

In section~\ref{loworder}, at $\eps \neq 0$,
we compute the vacuum energy density
and mass gap, according to the variationally
 improved perturbation theory of \cite{p1,p2,p3}
in the few lowest orders, adding and subtracting
a bare fermion mass, in closest
analogy with the procedure which converges so
 remarkably in the anharmonic oscillator
case. Not surprisingly, the procedure is in
general highly singular as
$\eps \rightarrow 0$, when the perturbative
order is fixed. However, for
$N \rightarrow \infty$, the variational procedure
 gives the exact answers
immediately when applied at first non trivial order.

In section~\ref{allorder}, using this remark and
the fact that for $N \rightarrow \infty$ the
theory reduces to cactus diagrams which can be
summed in closed (if implicit) form,
we show how to remove the singularity of the
procedure as $\eps \rightarrow 0$,
obtaining finite, \RG\ invariant expressions
for the vacuum energy and mass gap
of the massless theory in term of the physical,
 finite, \RG\ invariant mass
parameter of the effective massive theory, with
 respect to which we can then
optimize. The optimum, which gives the exact known
 answers, turns out to be
at zero, which is a physical consequence of the
fact that one is actually
solving the theory exactly. The procedure is
 sufficiently transparent to allow
for a generalization to arbitrary $N$,
which is done in the second paper.

\section{Low order calculations.} \label{loworder}

In $2+\eps$ dimensions, it is a straightforward
 exercise to compute
the vacuum graphs of Fig.~\ref{fig1}, which are
 the only ones which survives
in the large $N$ limit, using the Lagrangian
\begin{equation}
 {\cal L} = i \sum_{i=1}^{N} \bar{\psi_{i}}
\partial \hspace{-6pt} / \psi_{i}
           + m_{0} \sum_{i=1}^{N} \bar{\psi_{i}}
\psi_{i}   + {g_{0}^{2} \over 2} {( \sum_{i=1}^{N}
\bar{\psi_{i}} \psi_{i} )}^2 \hh ,\label{lagr}
\end{equation}
where $m_0$ and $g_0$ are bare parameters (in
the following, we shall suppress
the summation over the index $i$). We use the
bare parameters for several reasons.
First, from the remarkable results for the
anharmonic oscillator \cite{p2,p3}, we
may conjecture that our procedure converges
 to the exact answer \underline{for fixed
$\eps < 0$} for any $N$. Second, the use of
 renormalized parameters introduces
counterterms, which in effect mix perturbative
 orders and confuses matters as we are
already perturbing in $m_0$. It is thus
preferable to perform as much of the
calculation as possible with bare
 parameters, and only at the end relate them to
the renormalized ones via the \RG\
 through the standard procedure. The
 crucial
requirement is of course that the
final physical result remain finite for $\eps
\rightarrow 0$.

 We give the value
of the contribution of the graphs of Fig.~\ref{fig1}
to the energy density $E_0$ for all $N$ :
\begin{eqnarray}
E_{0} \, (m_{0}) & = &  \frac{m_{0}^{2+\eps} N \;
 \Gamma(-{\eps \over 2})}{(4 \pi)^{1+{\eps \over 2}}}
\; [ \; \frac{2}{2+\eps} \; + \; (2N-1) \, g_{0}^{2}
 \, m_{0}^{\eps} \; \frac{\Gamma(-{\eps \over 2})}
{(4 \pi)^{1+{\eps \over 2}}} \nonumber \\
 & & + \; (2N-1)^{2} \, g_{0}^{4} \, m_{0}^{2 \eps}
 \; (1+\eps) \; \frac{\Gamma^{2}(-{\eps \over 2})}
{(4 \pi)^{2+\eps}} \; + \; O(g_{0}^{6}) \; ] \nonumber \\
  & \equiv & E^{(0)} \, (m_{0}) \; + \; g_{0}^{2} \,
 E^{(1)} \, (m_{0}) \; +
     \; g_{0}^{4} \, E^{(2)} \, (m_{0}) \hh .
\end{eqnarray}
The straightforward application of the
variational procedure at lowest order calls
for the minimization with respect to $m_0$ of the function
\begin{equation}
E_{0}^{(1)} \, (m_{0}) = E^{(0)} \, (m_{0}) \; + \;
 g_{0}^{2} \, E^{(1)} \, (m_{0}) \; - \;
m_{0} \, \frac{\partial E^{(0)}\,(m_{0})}{\partial m_{0}}\hh .
\label{eord1}
\end{equation}
This gives the optimal value for the variational parameter
\begin{equation}
1=(2N-1)\;\frac{\Gamma(-{\eps\over 2})}{(4\pi)^{1+{\eps\over 2}}}
 \; g_{0}^{2}
 \, {m_{0}^{\eps}}_{(opt)} \hh ,
\label{mgap}
\end{equation}
and the corresponding optimum :
\begin{equation}
E_{0}^{(1)} =  {m_0}^{2+\eps}_{(opt)}  \; N \; \frac{
 \Gamma(-{\eps \over 2})}{(4 \pi)^{1+{\eps \over 2}}}
\; \frac{\eps}{2+\eps} \hh .
\label{egap}
\end{equation}
When $N$ goes to infinity, one can compute $E_0$
exactly using the effective potential
for the field $\sigma = \bar{\psi} \psi$. In this
 large-$N$ limit, one finds that
$E_{0}^{(1)}$ coincides with the exact $E_{0}$
and $m_0$ with the exact mass gap
for all values of $\eps$ . This fact has already
 been noted in Ref.~\cite{p2} and
corresponds to the fact that for $N = \infty$
the theory is a set of $N$ free massive
fermions, and the Hartree-Fock approximation
is exact. We notice that for $N = \infty$,
$\; g_{0}^{2} \,N \;$ fixed, $m_0$ and
$E_{0}^{(1)}$ as given by Eqs.~(\ref{mgap})~and~(\ref{egap})
are finite for $\eps \rightarrow 0$
provided that the renormalized coupling $g^2$,
given by
\begin{equation}
g_{0}^{2}=\frac{g^{2}\,{\mu}^{-\eps}}{1-\frac{(N-1) \,
 g^{2} }{\pi \, \eps} }  \hh ,
\label{barg}
\end{equation}
is kept fixed. However, $E^{(0)}$ and $g_{0}^{2}
 \, E^{(1)}$ are separately infinite.

The next order of the approximation calls
for the minimization with respect to $m_0$
of
\begin{eqnarray}
E_{0}^{(2)} \, (m_{0}) & = & E^{(0)} \, (m_{0})
 \; + \; g_{0}^{2} \, E^{(1)} \, (m_{0}) \; + \;
g_{0}^{4} \, E^{(2)} \, (m_{0}) \nonumber \\
 & & - \; m_{0} \, \frac{\partial E^{(0)} \,
 (m_{0}) }{\partial m_{0}} \; - \;
g_{0}^{2} \, m_{0} \, \frac{\partial E^{(1)}
 \, (m_{0}) }{\partial m_{0}} \nonumber \\
 & & + \; \frac{1}{2} \, m_{0}^{2} \,
\frac{{\partial}^{2} E^{(0)} \, (m_{0})
 }{\partial m_{0}^{2}} \hh .
\label{eord2}
\end{eqnarray}
Just as for the anharmonic oscillator,
and for the same reasons, $E_{0}^{(2)} \, (m_{0})$
has an extremum at the same value of $m_0$
 [Eq.~(\ref{mgap})] as $E_{0}^{(1)} \, (m_{0})$,
where it takes the same value [Eq.~(\ref{egap})],
 reflecting the fact that in the
large-$N$ limit, the variationally improved
perturbative treatment gives the exact answer
at each order.

Although the behaviour of the large-$N$ limit
 is satisfactory, this sample calculation
reveals two problems which must be solved
before on can hope to get interesting results
at finite $N$ or in more interesting theories.

The first problem can be seen at lowest
order, Eqs.~(\ref{mgap}),~(\ref{egap}) and (\ref{barg}),
which are true for all $N$: these equations
 do not give finite results for
$\eps \rightarrow 0$ at fixed finite $N$,
due to the mismatch between the coefficient
$(N-1)$ appearing in Eq.~(\ref{barg}) and
the coefficient $(N-1/2)$ appearing in Eq.~(\ref{mgap}):
the limits $N \rightarrow \infty$ and $\eps
 \rightarrow 0$ do not commute at this stage,
and it will be the essential content of this
 paper and the next one to make these two
limits commute, $i.e.$ to find how to reconcile
 in a general case a variational ansatz
with the fine tuning of renormalization,
materialized by Eq.~(\ref{barg}) and make the
physics continuous as $\eps$ crosses zero
 at fixed renormalized coupling constant
$g^2$ and mass scale $\mu$.

The other problem appears at second order
 in Eq.~(\ref{eord2}), and is again related
to the $\eps \rightarrow 0$ limit. In this
 equation one finds that the quantity
\begin{equation}
E^{(0)} \, (m_{0}) \; - \; m_{0} \,
 \frac{\partial E^{(0)} \, (m_{0}) }{\partial m_{0}} \;
+ \; \frac{1}{2} \, m_{0}^{2} \,
\frac{{\partial}^{2}E^{(0)}\,(m_{0})}{\partial m_{0}^{2}}
\end{equation}
is finite as $\eps \rightarrow 0$ although
 each term is separately divergent. Hence,
in some sense, the lowest order in perturbation
 theory becomes irrelevant relative to
the next two, each of order $\eps^{-1}$ (taking
 into account the fact that $g_{0}^{2}$
is of order $\eps$). This fact is independent
of the large-$N$ limit, and generalizes to
higher orders of the calculation: at any finite
 order, for any $N$, the last two orders
dominate all previous ones as $\eps \rightarrow
 0$. This is quite unpleasant, as one
would expect that all orders of perturbation
 theory contain at most a roughly equivalent
amount of information. This is even more
unpleasant in a perspective where one hopes
to obtain accurate results from a low-order
calculation!

These two pathologies of the $\eps \rightarrow
 0$ limit will be handled in the next
section in the large-$N$ limit.

Similar low-order calculations can be performed
 on the mass gap of the massive
Lagrangian~(\ref{lagr}). The contributing cactus
 diagrams up to order $g_{0}^{4}$
are on Fig.~\ref{fig2} and, for all $N$, they give simply
\begin{eqnarray}
m_{F} \, (m_0) & = &  m_{0} \; + \; (2N-1) \, g_{0}^{2}
 \, m_{0}^{1+\eps} \; \frac{\Gamma(-{\eps \over 2})}
{(4 \pi)^{1+{\eps \over 2}}} \nonumber \\
 & & + \; (2N-1)^{2} \, g_{0}^{4} \, m_{0}^{1+2 \eps}
 \; \frac{\Gamma^{2}(-{\eps \over 2})}
{(4 \pi)^{2+\eps}} \, (1+\eps) \; + \; O(g_{0}^{6}) \nonumber \\
 &\equiv & m_{F}^{(0)}\;+\;g_{0}^{2}\,m_{F}^{(1)}\;+
     \; g_{0}^{4} \, m_{F}^{(2)} \hh .
\label{mflow}
\end{eqnarray}
Perturbing at order 1 in both $g_{0}^{2}$ and $m_0$,
one has the mass gap
\begin{equation}
m_{F}^{(1)} \, (m_{0}) = m_{F}^{(0)} \; + \; g_{0}^{2}
 \, m_{F}^{(1)} \; - \;
m_{0} \, \frac{\partial m_{F}^{(0)}}{\partial m_{0}}
 \; = \; g_{0}^{2} \, m_{F}^{(1)} \hh .
\end{equation}

This has no non trivial extremum in $m_0$, but gives
 the exact large-$N$ mass gap
when the variational value of Eq.~(\ref{mgap})
is used for $m_0$.
Perturbing at order 2 in both $g_{0}^{2}$ and $m_0$ gives
\begin{equation}
m_{F}^{(2)} \, (m_{0}) = g_{0}^{2}
\, m_{F}^{(1)} \; - \; g_{0}^{2} \,
m_{0} \, \frac{\partial m_{F}^{(1)}}{\partial m_{0}}
 \; + \; g_{0}^{4} \, m_{F}^{(2)} \hh .
\end{equation}
Again, this gives the exact large-$N$ mass gap when
the value~(\ref{mgap}) for $m_0$
is used, but has a very singular behaviour if one
tries instead to extremize it
with respect to $m_0$. Indeed, in the limit $\eps
\rightarrow 0$, the term
$g_{0}^{4} \, m_{F}^{(2)}$ dominates, and it has
 no non trivial minimum.

These pathologies extend to higher finite orders
 of the large-$N$ limit:
plugging in the variational value (\ref{mgap})
for $m_0$ always gives the
exact large-$N$ mass gap, but for $\eps
 \rightarrow 0$, only the last order
of perturbation theory survives, excluding
 any useful direct variational
estimate of $m_F$; just as for the vacuum
energy, physics does not seem smooth
in the $\eps \rightarrow 0$ limit, and we
shall also solve this problem in
the next section.

\section{All order calculations and
 renormalization in the large-$N$ limit.}
 \label{allorder}

In this section, we fully exploit the fact
 that the large-$N$ limit can be solved in closed
(implicit) form and use further properties
of the variationally improved perturbation
theory discovered in one dimension (Ref.~\cite{p2})
 to make renormalization transparent
to our variational method, and the
physics continuous as $\eps$ goes to zero.

For the vacuum energy density, the large-$N$ limit
 amounts to retaining only the cactus
vacuum diagrams. For the Lagrangian density of
 Eq.~(\ref{lagr}), this can be evaluated
exactly and gives
\begin{eqnarray}
E(m_{0}) & = & \frac{- m_{0}^{2} \; N}{2 \, g_{0}^2 \;
[ \; 1 - \frac{(2N-1) \, g_{0}^{2} \, m_{F}^{\eps}
 \, \Gamma(-{\eps \over 2})}
{(4 \pi)^{1+{\eps \over 2}}} \; ] \;
( N - {1 \over 2} ) } \; + \;
\frac{m_{0}^{2}}{2\, g_{0}^2}\;\frac{N}{N-{1\over 2}}
   \nonumber \\
 & & + \; \frac{\eps}{2+\eps}\;\frac{N\,\Gamma(-{\eps
 \over 2})}{(4 \pi)^{1+{\eps \over 2}}}
\; m_{F}^{2+\eps} \hh ,
\end{eqnarray}
where $m_F$ is given by
\begin{equation}
m_F = \frac{ m_{0} }{1 - \frac{(2N-1) \, g_{0}^{2} \,
 m_{F}^{\eps} \, \Gamma(-{\eps \over 2})}
{(4 \pi)^{1+{\eps \over 2}}}}  \hh .
\label{mf}
\end{equation}

Physically, in the usual treatment of the model,
 $m_F$ is the mass gap, and is determined
from this last equation, which has a non trivial
 solution in the symmetric limit
$m_{0} \rightarrow 0$. In the present variational
 method, we must take into account the
perturbation in $m_{0}$ as shown in Eqs.~(\ref{eord1})
 and (\ref{eord2}). This is done rather
easily in terms of contour integrals. Define a function
 $f(x)$ by
\begin{equation}
f(x)=1-\frac{(2N-1)\, g_{0}^{2}\,\Gamma(-{\eps\over 2})}
{(4 \pi)^{1+{\eps \over 2}}} \; x \; m_{0}^{\eps} \;
 (1-x)^{\eps} \; f^{-\eps}  \hh .
\label{deff}
\end{equation}
This function has power series expansions in both
 $g_{0}^{2}$ and $x$,
beginning with 1.

The ${\textstyle n}^{\scriptstyle th}$ order of
 perturbation theory for the vacuum
energy density of the Lagrangian (\ref{lagr}) is then:
\begin{eqnarray}
E^{(n)} (m_{0}) & = & \frac{1}{2 i \pi} \; \oint
 \; \frac{dx}{x} \, x^{-n} \, (1-x) \;
[ \; - \, \frac{m_{0}^{2}}{2 \, g_{0}^2 \, f} \;
 + \; \frac{m_{0}^{2}}{2 \, g_{0}^2} \; ]
\; \frac{N}{N-{1 \over 2}} \nonumber \\
 & & \hspace{-50pt} + \; \frac{1}{2 i \pi} \;
 \oint \; \frac{dx}{x} \, x^{-n} \, (1-x)^{1+\eps} \;
 \frac{\eps}{2+\eps} \; \frac{N \, \Gamma(-{\eps
\over 2})\;m_{0}^{2+\eps}}{(4\pi)^{1+{\eps\over 2}}}
\; f^{-2-\eps} \hh ,
\label{eordn}
\end{eqnarray}
where the integration is counterclockwise on a small
circle around the origin.

It is of course straightforward to check that for
 $n=2$ one precisely reproduces
Eq.~(\ref{eord2}). It is also straightforward to
check that
\begin{equation}
\frac{\partial\,E^{(n)}(m_{0})}{\partial\, m_{0}} = 0
\end{equation}
at $m_0$ given by Eq.~(\ref{mgap}) and that the
corresponding extremum of $E^{(n)}$ is
independent of $n$, establishing in a few lines
that the variational result is independent
of the order of perturbation theory in the
large-$N$ limit. Indeed, for
\begin{equation}
1 = (2N-1) \; \frac{\Gamma(-{\eps \over 2})}{(4
\pi)^{1+{\eps \over 2}}} \; g_{0}^{2}
 \, {m_{0}^{\eps}} \hh ,
\end{equation}
one has simply
\begin{equation}
f(x) = 1 - x  \hh .
\end{equation}

Furthermore in general one may rewrite the
 definition of $f$, Eq.~(\ref{deff}) as
\begin{equation}
f(x) = (1-x) \; [ \; \frac{(4 \pi)^{1+{\eps
 \over 2}}}{(2N-1) \, g_{0}^{2} \,
 \Gamma(-{\eps \over 2})} \;]^{- \frac{1}{\eps}}
 \; (1-f)^{- \frac{1}{\eps}} \hh ,
\end{equation}
which shows that at fixed $\eps < 0$, $f$ can be
 expanded in a power series of $(1-x)$.
After a numerical exploration of the complex
 plane to make sure that no extra singularities
lie in the way, one may distort the integration
 contour in Eq.~(\ref{eordn}) to run clockwise
around the cut lying along the real positive
 axis and starting at $x=1$. Actually,
in Eq.~(\ref{eordn}), $x = 1$ is an isolated
pole, whose residue dominates its large $n$
behaviour; this immediately shows that in the
 large-$N$ limit, at fixed $\eps$ for any fixed
$m_0$, $E^{(n)} (m_{0})$ goes to the exact
 answer Eq.~(\ref{egap}), in the limit of
infinite order. This is reassuring: in the
 large $N$ case, which is well under control,
and where ordinary perturbation theory has
 a finite radius of convergence, the split
of Ref.~\cite{p1,p2} between free and
interacting terms is indeed independent of
$m_0$ in the limit of infinite order,
as intuition would suggest.

However one could go one step further,
 and, as already noticed in the case of the
anharmonic oscillator in Ref.~\cite{p2},
 extract more structure from the limit of
infinite order by rescaling $m_0$ with
the order $n$.

After distortion of the contour it is
clear that only the vicinity of $x = 1$
survives in the limit $n \rightarrow
\infty$, and one analyses this by the
change of variable
\begin{equation}
1 - x = \frac{v}{n} \hh .
\end{equation}
Rescaling $m_0$ by introducing $m_0 = m_{0}'
 \, n$, $m_{0}'$ kept fixed as $n$ goes to
infinity, one finds that $E^{(n)}$ in
Eq.~(\ref{eordn}) has a limit $E(m_{0}')$
given by
\begin{eqnarray}
E(m_{0}')  & = &  \frac{1}{2 i \pi} \;
 \oint \; v \, dv \, e^{v} \;
[ \; - \, \frac{{m_{0}'}^{2}}{2 \, g_{0}^2
 \, {f_{1}(v)} } \; + \;
\frac{{m_{0}'}^{2}}{2 \, g_{0}^2} \; ]
\; \frac{N}{N-{1 \over 2}} \nonumber \\
 & &  + \;  \frac{1}{2 i \pi} \;
\oint \; v \, dv \, e^{v} \;
\frac{ \eps \, N \, \Gamma(-{\eps \over 2})
 \; {m_{0}'}^{2+\eps} }{(2+\eps) \,
(4 \pi)^{1+{\eps \over 2}}}
\; {f_{1}(v)}^{-2-\eps} \hh ,
\end{eqnarray}
with $f_1$ defined simply by
\begin{equation}
f_{1}(v) = 1 - \frac{(2N-1) \, g_{0}^{2} \,
 \Gamma(-{\eps \over 2})}
{(4 \pi)^{1+{\eps \over 2}}} \; {(m_{0}' \,
 v)}^{\eps} \; {f_{1}(v)}^{-\eps} \hh ,
\label{deff1}
\end{equation}
and the $v$ integration contour lying
counterclockwise around
 the negative real axis in the cut $v$-plane.

The function $E(m_{0}')$ has the remarkable
property, not shared by $E^{(n)}(m_{0})$, to
allow for a smooth limit as $\eps \rightarrow
 0$. This follows from
the fact that Eq.~(\ref{deff1})
which defines $f_1$ is continuous as $\eps
 \rightarrow 0$ provided that $g_{0}^2$ and
$m_0$ follow their \RG\ behaviour. In this
limit, we obtain simply
\begin{equation}
E(m_{0}') = E ( m', g^{2} , \mu ) = \frac{{m'}^2}{2 i
 \pi} \; \oint \; \frac{ v \, dv \, e^{v} }{f_{2}^2} \;
\frac{ N }{4 \pi} \; ( \, 1 \; + \; \frac{2 \, f_2 \,
 \pi}{(N-{1 \over 2}) \, g^2} \, ) \hh ,
\label{eren}
\end{equation}
with
\begin{eqnarray}
f_{2} \, (v) & = & 1 \; + \; \frac{ g^{2} \, N }{\pi}
 \; \ln \frac{m' \, v }{ \mu \, f_2} \; - \;
\frac { N \, g^2 }{ 2 \pi } ( \, {\gamma}_{E} \, + \,
 \ln 4 \pi \, ) \nonumber \\
{\gamma}_{E} & = & 0.577215... \hh ,
\end{eqnarray}
where we have introduced the renormalized mass
\begin{equation}
m_{0}'=\frac{ m' }{1 -\frac{N\, g^{2}}{\pi\,\eps} } \hh .
\end{equation}

Eq.~(\ref{eren}) involves only renormalized quantities,
 and is renormalization group
invariant in the sense of the effective massive theory:
\begin{equation}
[ \; \mu \, {\partial}_{\mu} \; + \beta (g) \, g \,
 \frac{\partial}{\partial \, g} \, - \;
\gamma (g) \, m \, \frac{\partial}{\partial \, m} \;
 ] \; E ( m', g^{2} , \mu ) = 0 \hh ,
\end{equation}
with
\begin{equation}
\beta (g) =  - \, \frac{g^2 \, N}{2\pi} \hh ,
\hh \gamma (g) = \frac{g^2 \, N}{\pi} \hh .
\end{equation}
This follows trivially from the fact that $E ( m',
g^{2} , \mu )$ depends only on the bare
parameters $m'_0$ and $g_{0}^2$. What is much less
trivial is that we have arrived at a
finite quantity for $\eps \rightarrow 0$ :
$E^{(n)}(m_0)$ in Eq.~({\ref{eordn}) is also
\RG\ invariant but not finite in the limit
$\eps \rightarrow 0$, while $E(m'_0)$ is both
invariant and finite.

One can go one step further by defining the
 dimensionless parameter
\begin{equation}
m''=\frac{m'\,\pi}{N\,g^2\,\mu\, e^{-\frac{\pi}{N g^2}}}\,
\end{equation}
and a function $f_{3}$ by
\begin{equation}
f_{3} =  \ln ( \, m'' \, v \, ) \; - \; \ln f_{3}
\label{def} \hh .
\end{equation}
$m''$ is a finite pure number invariant by the
\RG\ of the effective theory.

Studying the function $E(m'')$ poses no problem.
 One finds that the function
$f_{3}$ defined by Eq.~(\ref{def}) behaves as
 $ \ln m'' + O ( \ln \ln m'' )$
as $m''$ goes to infinity, a typical
\RG\-improved high energy behaviour.
Furthermore, $f_3$ can be expanded in power
 series of $v$ around $v = 0$, with
radius of convergence $1/e$, and with a cut
 extending from $-1/e$ ( a square
root branch point ) to $- \infty$. The
behaviour of  $E(m'')$ for large $m''$
can be obtained by standard methods of
contour integrals, and is of the form
\begin{equation}
E(m'') \hspace{10pt} {\sim}_{ \hspace{-20pt}
 \raisebox{-1.ex}{${\scriptstyle m''
\rightarrow\infty }$ } }
 \frac{{m''}^2}{\ln m''} \hh .
\end{equation}
This is the perturbative regime. For $m''
 \rightarrow 0$, $E(m'')$ converges exponentially
to
\begin{equation}
E(0)=-\frac{N}{4 \pi}\,\mu^2\,
e^{-\frac{2\pi}{N g^2}}\hh ,
\end{equation}
which is the only real extremum. Plotting
 the curve $E(m'')$ poses no problem
with a workstation, and the result can be
 seen on Fig.~\ref{fig3}. Notice its
remarkably smooth behaviour.

The extremum at $m'' = 0$ is the exact
value derived from the original calculation using the
effective potential of the field $\sigma =
 \bar{\psi} \psi$. This fact is consistent
with the rescaling $m_{0} = {m'}_{0} \, n$
 as the order goes to infinity, and the
fixed position of the extremum of $E^{(n)}$
 given by Eq.~(\ref{mgap}).

The same procedure can be applied to the
calculation of the mass gap. From
Eq.~(\ref{mf}), the ${\textstyle n}^
{\scriptstyle th}$ order of perturbation
theory for the mass gap is
\begin{equation}
m_{F}^{(n)} =  \frac{1}{2 i \pi} \;
\oint \; \frac{dx}{x} \, x^{-n} \;
\frac{m_0}{f} \hh .
\end{equation}
This reproduces the low order calculation
 (\ref{mflow}); when $m_0$ is the value
of Eq.~(\ref{mgap}) which extremises $E^{(n)}$,
 it is immediate that $m_{F}^{(n)} = m_0$ ;
one may also go to infinite order, following
the same rescaling of $1-x$ and $m_0$
as for the vacuum energy density. In this
infinite order limit,one obtains
\begin{equation}
m_{F} ( m', g^{2} , \mu ) = \frac{m'}{2 i \pi}
 \; \oint \; \frac{dv \, e^{v} }{f_{2}} \hh ,
\end{equation}
a formula where the $\eps \rightarrow 0$ limit
 has been taken like in the case
of $E ( m', g^{2} , \mu )$, and contains the
variational parameter $m'$ in a finite
\RG\ invariant way.

$m_{F} ( m', g^{2} , \mu )$ is actually a
function of the
renormalization invariant parameter $m''$
 only; $m_{F} ( m'' ) $ can be analyzed like $E(m'')$,
approaching exponentially the exact value
 at $m'' = 0$, its only real extremum.
It is plotted in Fig.~\ref{fig4}.

\section{Conclusion.}

In this paper, with the Gross-Neveu model in
 the large $N$ limit as example, we have
studied some pathologies of the variational
method applied to a renormalizable quantum field
theory. On the two examples of the vacuum
energy density and the mass gap,
we have shown how a resummation to all orders
 can make these pathologies disappear,
and make a variational method perfectly
compatible with renormalization, {\em i.e.} one  may
replace the original theory ( the massless
model ) by another one ( a massive model ),
varying the parameters while keeping physics
 continuous as the space-time dimension
is varied around its critical value. In the
next paper, we extend
the procedure to the finite-$N$ case, where
the mathematics of the \RG\ and
the Feynman diagram stuctures are more generic.

\section{Acknowlegments}

One of us (A. N.) is grateful for their
hospitality to the
Lawrence Berkeley Laboratory,
where this work was supported in part by the Director,
Office of Energy research, Office of High Energy and
Nuclear Physics, Division of High Energy Physics of
the U.S. Department of Energy under Contract
DE-AC03-76SF00098 and to the Physics Department,
University of California,
Berkeley, where this work was completed with partial
support from National Science Foundation
grant PHY-90-21139.
(C.A.) is grateful to the theory group of Imperial
College for their  hospitality
and acknowledges financial
 support from the Minist\`{e}re de
la Recherche et de la Technologie, and from E.E.C. grant
No. ERBCHBICT941235.

\newpage
\centerline{\large \bf Figure Captions.}

\vspace{20pt}

\noindent
Figure 1. Vacuum energy graphs in the large-$N$ limit.

\noindent
Figure 2. Mass gap graphs in the large-$N$ limit.

\noindent
Figure 3. Variational vacuum energy $E(m'')/E(0)$
 as a function of $m''$.

\noindent
Figure 4. Mass gap $m_F(m'')/m_F(0)$ as a function
 of the variational parameter $m''$.

\newpage

\begin{figure}[t]
\epsfxsize=8truecm
\epsfysize=4truecm
\centerline{\epsffile{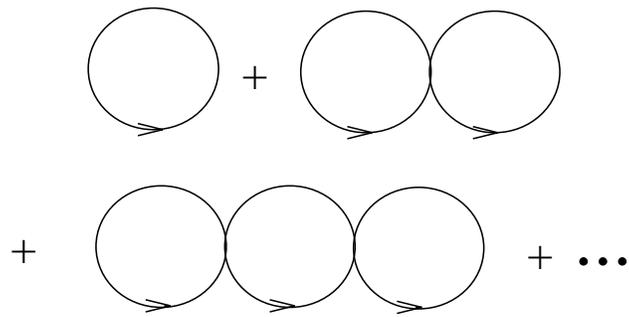}}
\vspace{1.cm}
\caption{Vacuum energy graphs in the large-$N$ limit.}
\label{fig1}
\end{figure}

\vspace{60pt}

\begin{figure}[b]
\epsfxsize=12truecm
\epsfysize=3truecm
\centerline{\epsffile{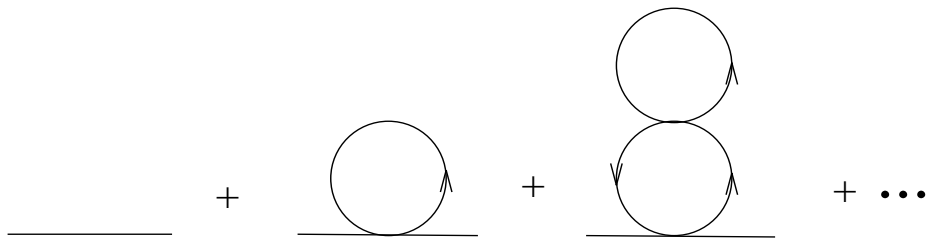}}
\vspace{1.cm}
\caption{Mass gap graphs in the large-$N$ limit.}
\label{fig2}
\end{figure}

\begin{figure}[p]
\epsfxsize=16truecm
\epsfysize=16truecm
\centerline{\epsffile{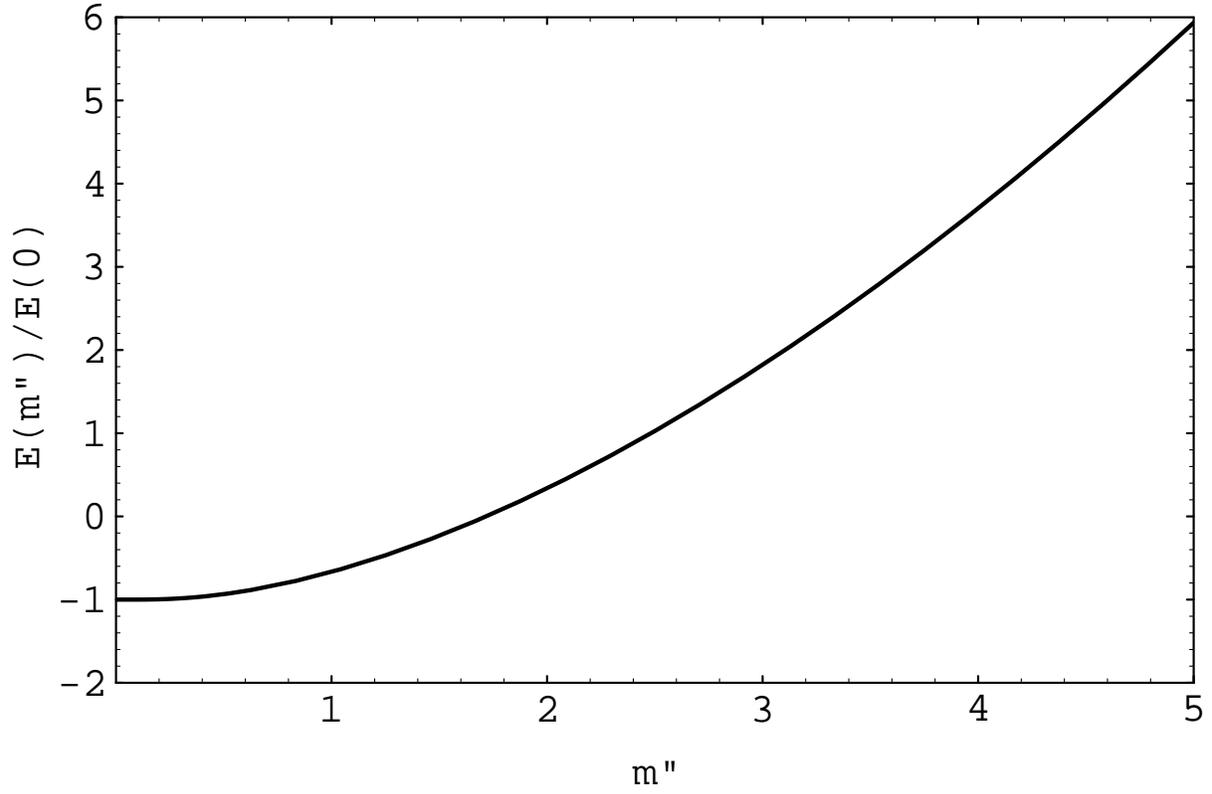}}
\caption{Variational vacuum energy $E(m'')/E(0)$
as a function of $m''$.}
\label{fig3}
\end{figure}

\begin{figure}[p]
\epsfxsize=16truecm
\epsfysize=16truecm
\centerline{\epsffile{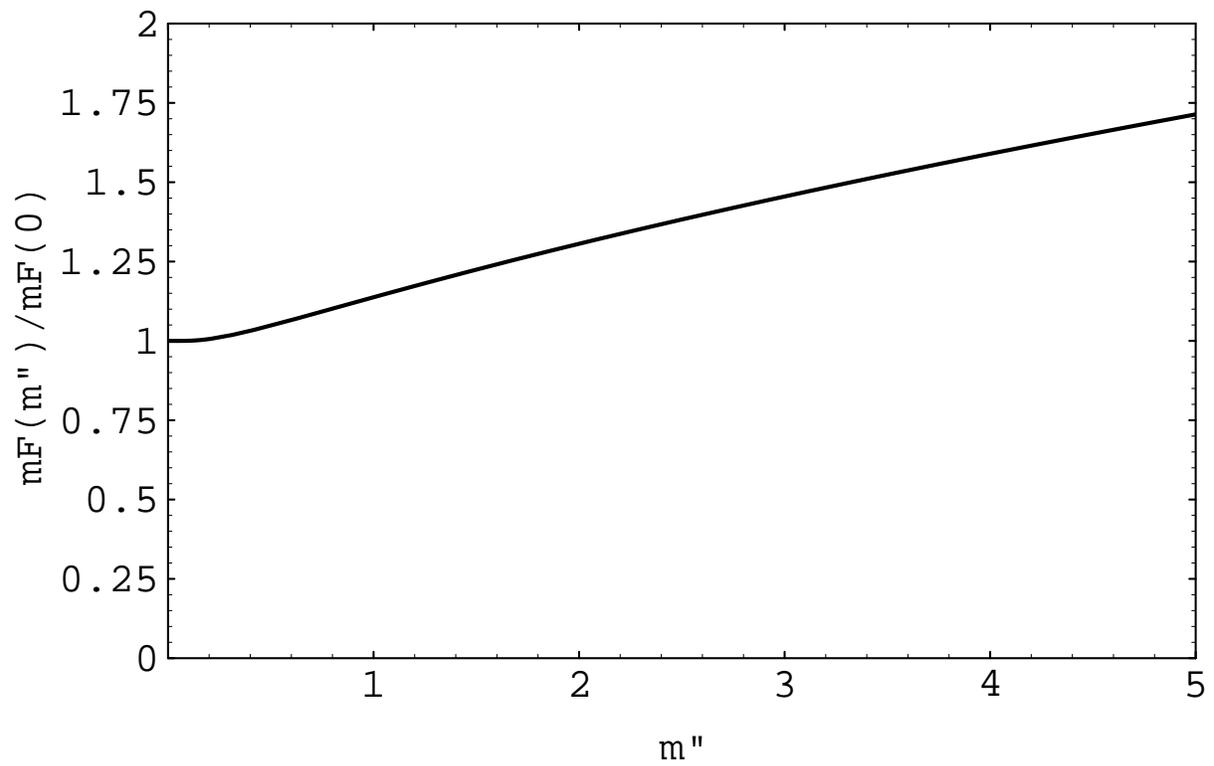}}
\caption{Mass gap $m_F(m'')/m_F(0)$ as a function
of the variational parameter $m''$.}
\label{fig4}
\end{figure}

\end{document}